\documentclass[twocolumn]{aastex63}

\usepackage[T1]{fontenc}
\usepackage{bm}        
\usepackage{amssymb, amsmath}   
\usepackage{cancel}
\usepackage{color}

\DeclareMathAlphabet{\mathsfit}{\encodingdefault}{\sfdefault}{m}{sl}
\SetMathAlphabet{\mathsfit}{bold}{\encodingdefault}{\sfdefault}{bx}{sl}
\newcommand{\vect}[1]{\bm{#1}}

\DeclareMathOperator{\sech}{sech}
\shorttitle{Multi-Wavelength Signatures from Reconnection}
\shortauthors{Zhang et al.}

\begin{document}

\defcitealias{Zhang2020}{Paper I}

\title{Radiation and Polarization Signatures from Magnetic Reconnection in Relativistic Jets--II. Connection with $\gamma$-rays}

\correspondingauthor{Haocheng Zhang}
\email{astrophyszhc@hotmail.com}

\author[0000-0001-9826-1759]{Haocheng Zhang}
\affiliation{New Mexico Consortium \\
Los Alamos, NM 87544, USA}
\affiliation{Department of Physics and Astronomy \\
Purdue University \\
West Lafayette, IN 47907, USA}

\author[0000-0001-5278-8029]{Xiaocan Li}
\affiliation{Dartmouth College\\
Hanover, NH 03750, USA}

\author[0000-0003-1503-2446]{Dimitrios Giannios}
\affiliation{Department of Physics and Astronomy \\
Purdue University \\
West Lafayette, IN 47907, USA}

\author[0000-0003-4315-3755]{Fan Guo}
\affiliation{Theoretical Division \\
Los Alamos National Lab \\
Los Alamos, NM 87545, USA}
\affiliation{New Mexico Consortium \\
Los Alamos, NM 87544, USA}

\author[0000-0002-4024-3280]{Hannes Thiersen}
\affiliation{Centre for Space Research \\
North-West University \\
Potchefstroom, 2520, South Africa}

\author[0000-0002-8434-5692]{Markus B\"ottcher}
\affiliation{Centre for Space Research \\
North-West University \\
Potchefstroom, 2520, South Africa}

\author[0000-0002-9854-1432]{Tiffany Lewis}
\affiliation{NASA Postdoctoral Program Fellow}
\affiliation{NASA Goddard Space Flight Center\\
Greenbelt, MD 20771, USA}

\author[0000-0002-4188-627X]{Tonia Venters}
\affiliation{Astrophysics Science Division \\
NASA Goddard Space Flight Center\\
Greenbelt, MD 20771, USA}

\begin{abstract}
It is commonly believed that blazar jets are relativistic magnetized plasma outflows from supermassive black holes. One key question is how the jets dissipate magnetic energy to accelerate particles and drive powerful multi-wavelength flares. Relativistic magnetic reconnection has been proposed as the primary plasma physical process in the blazar emission region. Recent numerical simulations have shown strong acceleration of nonthermal particles that may lead to multi-wavelength flares. Nevertheless, previous works have not directly evaluated $\gamma$-ray signatures from first-principle simulations. In this paper, we employ combined particle-in-cell and polarized radiation transfer simulations to study multi-wavelength radiation and optical polarization signatures under the leptonic scenario from relativistic magnetic reconnection. We find harder-when-brighter trends in optical and {\it Fermi-LAT} $\gamma$-ray bands as well as closely correlated optical and $\gamma$-ray flares. The optical polarization angle swings are also accompanied by $\gamma$-ray flares with trivial time delays. Intriguingly, we find highly variable synchrotron self Compton signatures due to inhomogeneous particle distributions during plasmoid mergers. This feature may result in fast $\gamma$-ray flares or orphan $\gamma$-ray flares under the leptonic scenario, complementary to the frequently considered mini-jet scenario. It may also infer neutrino emission with low secondary synchrotron flux under the hadronic scenario, if plasmoid mergers can accelerate protons to very high energy.
\end{abstract}

\keywords{galaxies: jets --- radiation mechanisms: non-thermal --- magnetic reconnection --- polarization}


\section{Introduction} \label{sec:intro}

Blazars are relativistic plasma jets from fast accreting supermassive black holes that point very close to our line of sight.  They are the most numerous extragalactic $\gamma$-ray sources and among the most powerful astrophysical phenomena in the Universe \citep{Fermi2019}. The blazar emission is dominated by nonthermal radiation processes with two spectral components \citep[see][for a recent review]{Boettcher2019}. The low-energy component from radio to optical, in some cases up to X-rays, is dominated by synchrotron emission from ultra-relativistic electrons. The typical polarization degrees (PD) observed from radio to optical bands are consistent with electron synchrotron in a partially ordered magnetic field, supporting this idea \citep{Pushkarev2005,Zhang2015}. The high-energy component from X-rays to $\gamma$-rays is often interpreted as inverse Compton scattering by the same electrons that make the synchrotron component \citep[often referred to as the leptonic model, see][]{Marscher1985,Dermer1992,Sikora1994}. The seed photons can be the low-energy synchrotron component itself, in which case it is called the synchrotron self Compton (SSC), or come from external photon fields such as the emission from broad line region and dusty torus, referred to as the external Compton (EC). Alternatively, hadronic emission processes, including proton synchrotron, photomeson, and subsequent cascades, can contribute to the high-energy component \citep[referred to as the hadronic model, see][]{Mannheim1993,Mucke2003,Boettcher2013}. In particular, the recent detection of a very high energy neutrino is potentially correlated with a blazar flare, strongly suggesting the presence of hadronic processes in blazars \citep{Icecube2018,Keivani2018,Gao2019,Cerruti2019,Reimer2019,Zhang2019}. It has been suggested that X-ray to MeV $\gamma$-ray polarimetry can independently diagnose the contribution of hadronic processes in the high-energy spectral component \citep{Zhang2013,Zhang2016,Zhang2017b,Paliya2018}, which may be detected by the upcoming {\it IXPE}\footnote{\url{https://ixpe.msfc.nasa.gov/}} as well as future MeV $\gamma$-ray telescopes such as {\it AMEGO} \citep{McEnery19,Rani2019}.

The multi-wavelength blazar emission can be highly variable. In some extreme events, the GeV to TeV $\gamma$-rays can flare within a few minutes, implying efficient particle acceleration in very localized regions in the blazar jet \citep{Aharonian2007,Albert2007,Ackermann2016}. Observations have shown that the low- and high-energy components often flare together, which is consistent with a leptonic origin of the blazar emission \citep{Chatterjee2012,Liodakis2019}. Moreover, the optical polarization signatures appear variable as well \citep[][and see \citealt{Zhang2019b} for a recent review]{Smith2009,Ikejiri2011}. Especially during flares, the optical PD and polarization angle (PA) can change considerably. Very interestingly, observations have seen large optical PA rotations during blazar flares, suggesting that the magnetic field morphology may undergo significant changes \citep{Marscher2008,Larionov2013,Blinov2015,Chandra2015}. Statistical studies by the RoboPol project\footnote{\url{https://robopol.physics.uoc.gr/}} have revealed that the optical PA swings are correlated to {\it Fermi-LAT} $\gamma$-ray flares, and that the correlation is unlikely to be purely stochastic \citep{Angelakis2016,Blinov2016,Kiehlmann2017,Blinov2018}. It is therefore important to uncover the physical driver that can simultaneously explain the correlated multi-wavelength radiation and polarization signatures.

Relativistic magnetic reconnection is a promising physical mechanism to model blazar flares. It is a plasma physical process where oppositely directed magnetic field lines come close to each other, rearrange their magnetic topology and release a considerable amount of magnetic energy \citep[see][for a review on recent progress]{Guo2020}. The magnetic dissipation process can efficiently accelerate particles to nonthermal distributions if the environment is magnetically dominated. Recent numerical simulations have shown that reconnection can accelerate both electrons and protons into power-law spectra \citep{Guo2014,Guo2015,Guo2016,Sironi2014,Werner2018,Werner2021,li2018,Li2019,kilian2020,Sironi2015,Petropoulou2019}. The spectral indices depend on the physical parameters of the reconnection region, especially the magnetization factor ($\sigma$, the ratio between magnetic energy and enthalpy densities). Furthermore, particle acceleration in reconnection can be very fast if $\sigma>1$, and the reconnection outflows can also be relativistic, making reconnection a very natural scenario for the very fast $\gamma$-ray variability observed in blazars \citep{Giannios2009,Christie2020,Sironi2016,Liu15}.

Several works have explored the blazar spectra and light curves resulting from relativistic magnetic reconnection in the blazar emission environment \citep{Deng2016,Petropoulou2016,Yuan2016,Kagan2016,Mehlhaff2020,Christie2020}. Their results are consistent with typical observations. Very interestingly, \citet{Zhang2018} have revealed that reconnection can lead to large optical PA swings during blazar flares, which has been later verified by \citet{Hosking2020}. However, those previous works are mostly case-by-case studies, lacking exploration of systematic patterns and correlations between multi-wavelength signatures. \citet{Zhang2020} \citepalias[hereafter][]{Zhang2020} have made the first attempt to systematically study blazar radiation and polarization signatures from relativistic magnetic reconnection. It has hinted at several observable patterns, including the harder-when-brighter trend in the synchrotron component and correlations between polarization signatures with observational bands. In this paper, we will extend our previous work to the high-energy spectral component assuming a leptonic origin. We will examine the multi-wavelength variability and time delays between flares in low- and high-energy bands. Additionally, we will study the connection between optical polarization signatures, in particular the optical PA swings, and $\gamma$-ray flares. Furthermore, we will survey the effects of synchrotron and Compton scattering cooling on the plasma dynamics and radiation signatures arising from reconnection. Section \ref{sec:setup} describes our simulation setup, Section \ref{sec:results} shows the multi-wavelength signatures from our simulations with an emphasis on the SSC emission, Section \ref{sec:observation} discusses the implications for observations, and Section \ref{sec:discussion} summarizes and discusses our results.

\section{Simulation Setup \label{sec:setup}}

We assume that reconnection happens in a considerably magnetized blazar emission environment from a pre-existing current sheet. Such structures may be present in a kink-unstable emission region or a striped jet configuration \citep{Begelman1998,Bodo2021,Zhang2017,Giannios2019,Zhang2021,Giannios2006}. Our simulations mostly follow the setup in \citetalias{Zhang2020}, except for the addition of Compton scattering. In the following, we will briefly describe our simulation setup with an emphasis on the treatment of Compton scattering.

\subsection{PIC Setup}

We perform 2D PIC simulations in the $x$-$z$ plane using the \texttt{VPIC} code \citep{Bowers2008}, with the length in the $x$-direction $L_x=2L$ and that in the $z$-direction $L_z=L$. Simulations start from a force-free current sheet, with the magnetic field $\vect{B}=B_0\tanh(z/\lambda)\hat{x}+B_0\sqrt{\sech^2(z/\lambda)+B_g^2/B_0^2}\hat{y}$, where $B_g=0.2B_0$ is the strength of the guide field, the component perpendicular to the anti-parallel components $B_0$. We assume an electron-proton plasma with realistic mass ratio $m_i/m_e=1836$. We set the half-thickness of the current sheet to be $\lambda=0.6\sqrt{\sigma_e}d_{e0}$, where $d_{e0}=c/\omega_{pe0}$ is the nonrelativistic electron inertial length, $\omega_{pe0}=\sqrt{4\pi n_ee^2/m_e}$ is the nonrelativistic electron plasma frequency, and $\sigma_e=B_0^2/(4\pi n_em_ec^2)$ is the cold electron magnetization parameter \citep{Sironi2014,Guo2014}. Initially both electron and proton distributions are Maxwell-J\"uttner distributions with uniform density $n_e=n_i$ and temperature $T_e=T_i=100 m_e c^2$. Therefore, the upstream electron inertial length is $d_e=\sqrt{1+3T_e/(2m_ec^2)}d_{e0}\sim 12.2 d_{e0}$. The simulation box length scale is set to be $L=8\times 10^3d_{e0}$. Therefore, the simulation box size is $16000d_{e0}\times 8000d_{e0}$, or $1311d_e\times 653d_e$ in units of upstream electron inertial length. The resolution is set to be $4096\times 2048$, so that the cell sizes $\Delta x= \Delta z\sim 0.32 d_e$ is enough to resolve the upstream electron inertial length. We set the initial $\sigma_e=4\times 10^4$, so that the total magnetization factor $\sigma_0\sim \sigma_e m_e/m_i \sim 22$.

We implement a radiation reaction force to mimic the cooling effect in blazars, which can be considered as a continuous friction force for relativistic electrons \citep[non-relativistic terms are ignored, see][]{Cerutti2012,Cerutti2013,Cerutti2017},
\begin{align*}
  \vect{g}
  = -\frac{2}{3}r_e^2\gamma\left[\left(\vect{E}+
  \frac{\vect{u}\times\vect{B}}{\gamma}\right)^2 -
  \left(\frac{\vect{u}\cdot\vect{E}}{\gamma}\right)^2\right]\vect{u}-\frac{4}{3}\sigma_T\gamma \mathcal{U}_{\star}\vect{u},
\end{align*}
where $\vect{u}=\gamma \vect{v}/c$ is the four-velocity, $r_e=e^2/m_ec^2$ is the classical radius of the electron, and $\mathcal{U}_{\star}$ is the photon energy density. However, due to the very small scale of the PIC simulation, the typical blazar cooling parameters have trivial effects on PIC scales. Here we scale up the above cooling force by multiplying a factor so that the cooling break of the particle spectrum happens at $\gamma_c\sim 10^4$. As we will see in the next section, this will lead to a cooling break at $\sim 10^4$ in about one light crossing time. With this setup, we estimate that the so-called radiation-reaction (burn-off) limit, where the cooling becomes comparable with the Lorentz force, is at $\gamma_{rad}\sim 6 \times 10^4$ \citep{Uzdensky2011}. Although this burn-off limit is much lower than the typical blazar emission environment due to the normalization of $\vect{g}$, it does not affect our radiation signatures. This is because the highest-energy $\gamma$-rays that we are interested in only extend to $\sim 10~\rm{GeV}$, which is dominated by electrons at $\gamma_c \sim 10^4$. For simplicity, we only consider a stationary and uniform photon field $\mathcal{U}_{\star}$ that represents the EC. Therefore, the radiation reaction force consists of the local cooling due to synchrotron and the global cooling due to Compton scattering. In principle, the cooling due to SSC can be important, which naturally varies in space and time given the inhomogeneous and fast-evolving synchrotron photon field. However, as we will see in the following, mostly the SSC is strong only in very localized regions, making SSC cooling a local effect, just like synchrotron cooling. We will also see in Section \ref{sec:cooling} that the synchrotron and Compton scattering show similar cooling effects on particles (this is expected, as both are proportional to $\gamma^2$ if the Compton scattering is in the Thomson regime), thus only the total cooling rate is important to the particle evolution. For simplicity, here we do not explicitly consider the SSC cooling in the PIC setup.

\subsection{Radiation Transfer Setup}

Some previous works suggest that for magnetic reconnection in either a kink-unstable jet or a striped jet, the blazar emission region is likely around 1 pc from the central engine \citep{Dong2020,Zhang2021}. In this situation, the external photon field for the EC component may include contributions from the accretion disk, broad line region, molecular cloud, and dusty torus. For simplicity, here we consider a black body spectrum at 5000 K to mimic the effect of the above combined thermal emission components. We assume a typical bulk Lorentz factor of 10 for the reconnection region. Therefore, the external photon spectrum peaks at $\sim 12~\rm{eV}$ in the comoving frame of our simulation. Its photon energy density is considered as a free parameter for our simulation, which is essentially $\mathcal{U}_{\star}$ in the PIC simulation setup.

Here we show how the above simulation parameters translate to physical units. Leptonic blazar spectral fitting studies often find the magnetic field in the emission region is somewhere around $B\sim 0.1~\rm{G}$ \citep{Boettcher2013,Paliya2018}. Provided our simulation parameters, the low- and high-end of the particle spectrum in the emission region will be at $\gamma_1\sim 10^2$ and $\gamma_2\sim 10^4$, respectively, which are typical with spectral fitting studies. With a bulk Lorentz factor of ten, the above parameters imply the low-energy spectral component peaks at the optical band, while the high-energy component peaks around GeV. Provided the above magnetization factor, we can find that the electron number density is $n_e\sim 0.02 \rm{cm^{-3}}$. The size of the simulation box is then $L_x\sim 6\times 10^{10}~\rm{cm}$ and $L_z \sim 3\times 10^{10}~\rm{cm}$. Although this is much smaller than the typical blazar emission region size at $\sim 10^{16}~\rm{cm}$, it is much larger than the typical kinetic length scale of the highest-energy electrons in the simulation (for instance, gyroradius $r_g\sim 1.5\times 10^8~\rm{cm}$). As suggested in previous works, the simulation results do not appear to depend on the box size as long as it is much larger than the particle kinetic length scales \citep{Sironi2016,Petropoulou2016,Petropoulou2019,Zhang2020,Christie2019}.

We describe our radiation transfer setup as follows. We fix our line of sight in the guide field direction ($y$-axis) in the comoving frame of the simulation domain, similar to our previous works. We assume that the simulation domain is moving in the $z$-direction with a bulk Lorentz factor $\Gamma=10$. In this way, we have the relativistic Doppler factor $\delta=\Gamma=10$. The PIC simulation results are reduced every $16\times 16$ PIC cells into one radiation transfer cell to obtain adequate statistics for the particle spectrum. For each radiation transfer cell, we divide the particle kinetic energy $(\gamma-1)m_ec^2$ into 100 steps in the logarithmic scale between $10^{-4}m_ec^2$ to $10^6m_ec^2$, and obtain the particle spectrum by counting the number of particles in each step. We obtain the magnetic field by directly averaging the magnetic field in the $16\times 16$ PIC cell. The spatially resolved, time-dependent magnetic field and particle distributions are fed into the \texttt{3DPol} code developed by \citet{Zhang2014} to calculate the synchrotron emission. This code calculates the Stokes parameters of the synchrotron emission, which represent the total and polarized emission from each radiation transfer cell, then uses a ray-tracing method to trace the emission to the plane of the sky. There it adds up all the synchrotron emission that arrives in the same time step within the same spatially resolved cell on the plane of the sky. In the end, the code outputs the total spectra, light curves, polarization signatures from the entire simulation domains, as well as snapshots of the polarized emission map.

For the high-energy spectral component, we consider both SSC and EC contributions (even though the SSC cooling is not explicitly included in the PIC setup). Although EC is often dominating in the GeV $\gamma$-rays, SSC can still make considerable contribution \citep{Boettcher2013}, and it dominates the $\gamma$-ray emission for BL Lac objects. Since the particle spectra are given by the PIC simulation and the synchrotron photon field evolution is given by the radiation transfer, we can calculate the SSC spectra and light curves easily. \citet{Christie2020} have shown that if the plasmoid motion in the simulation domain is relativistic, the local Doppler boosting from these plasmoids (often referred to as the ``mini-jet'') can significantly boost the SSC emission. Here plasmoids refer to quasi-circular magnetic structures with a concentration of nonthermal particles that show up in the reconnection region. In our simulations, owing to the presence of the nontrivial guide field and periodic boundary conditions, we do not see any bulk relativistic motion of plasmoids in the outflow direction during the reconnection. Therefore, we do not consider the local Doppler boosting in our radiation transfer simulations. However, the synchrotron photon field in the reconnection layer is highly inhomogeneous. Then it is possible that the synchrotron photon field in certain locations in the simulation domain is not dominated by the local synchrotron emission, but by synchrotron from some large plasmoids or plasmoid merger events. Therefore, unlike the EC contribution, where the external photon field is often considered as a uniform photon field, the SSC emission may be delayed due to seed photons having to first travel from the location of larger plasmoid or plasmoid merger event to the local Compton scattering region, and then travel to the observer once scattered. This so-called ``internal light crossing time delay'', well described in \citet{Chen2012,Chen2014}, can lead to a small time delay between the SSC and EC emission. However, as we will see in the following, the SSC from magnetic reconnection is only strong in very localized regions with enhanced particle acceleration. Consequently, we can safely ignore the internal light crossing effects and only consider the local synchrotron photon field in each radiation transfer cell for the SSC. This is done through a Compton scattering module that works similarly as the \texttt{3DPol} code, but without the polarization dependence. This is because the PD due to SSC and EC is generally very low in the blazar emission region, which is unlikely to be observed \citep{Bonometto1970,Paliya2018,Zhang2019,Dreyer2021}. To summarize, we consider two seed photon fields for the Compton scattering in the radiation transfer simulation: an inhomogeneous and fast-evolving synchrotron photon field for the SSC, and a uniform and stationary black body photon field for the EC.

\section{Multi-Wavelength Signatures from Magnetic Reconnection \label{sec:results}}

Under the leptonic emission model, we find that the optical and $\gamma$-ray light curves from magnetic reconnection are highly correlated during blazar flares. In particular, the optical PA swings are physically connected to $\gamma$-ray flares. Very interestingly, the SSC light curves are much more variable than those of the EC. Specifically, the SSC can exhibit very strong and fast flares within $\lesssim 0.1~\tau_{lc}$, where $\tau_{lc}$ is the light crossing time scale of the reconnection region, owing to the extremely inhomogeneous particle acceleration during plasmoid mergers. In this section, we present the multi-wavelength signatures from relativistic magnetic reconnection. Our default setup considers comparable cooling due to synchrotron and Compton scattering, i.e., magnetic energy density is comparable to the photon energy density, $u_B\sim \mathcal{U}_{\star}$. Then we will present a parameter study about the ratio between the two cooling effects.

\subsection{Optical and Gamma-Ray Signatures \label{sec:opticalgamma}}

\begin{figure}
\centering
\includegraphics[width=\linewidth]{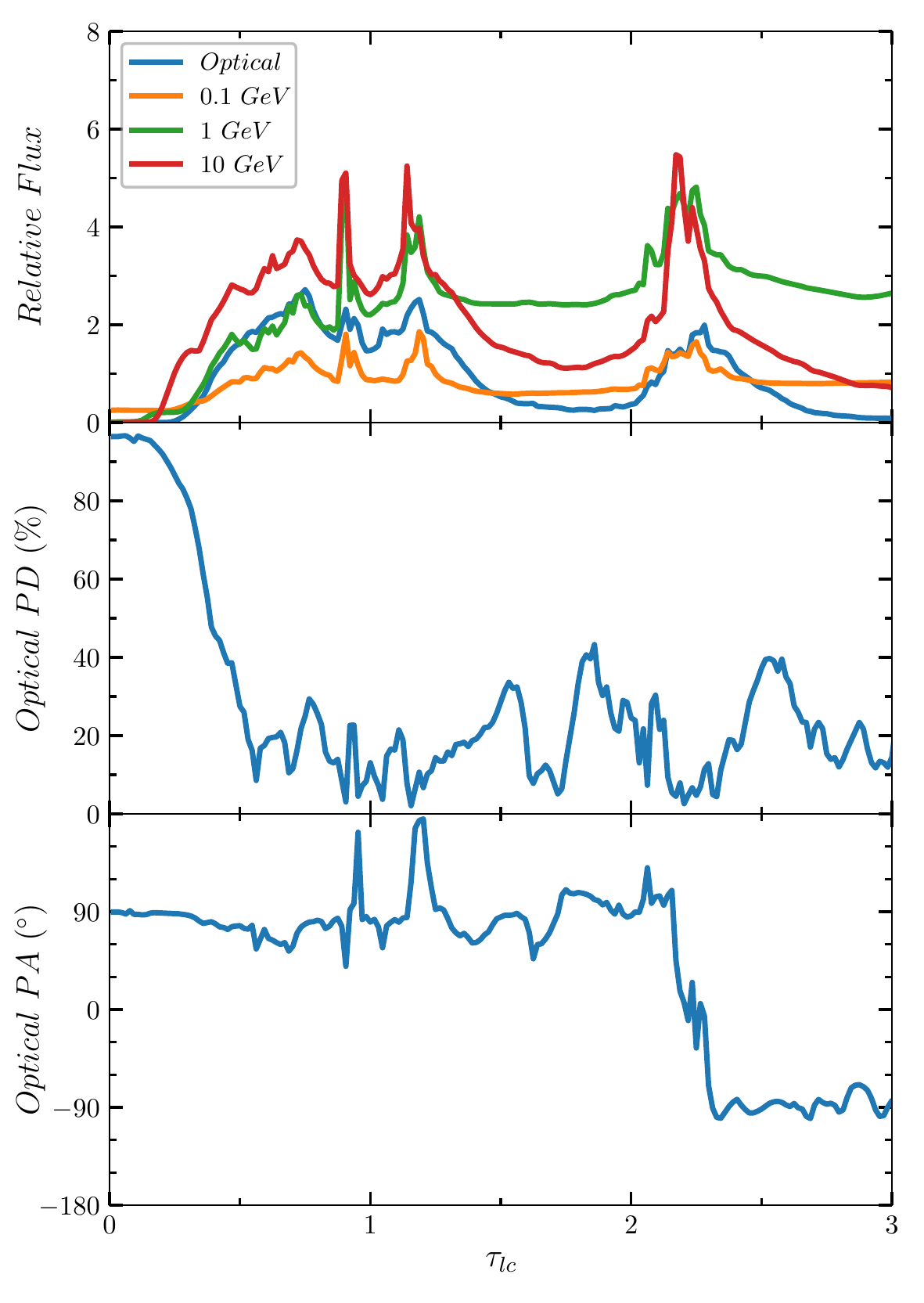}
\caption{Time-dependent signatures from the default simulation. The upper panel shows the light curves in the optical, $0.1~\rm{GeV}$, $1~\rm{GeV}$, and $10~\rm{GeV}$ $\gamma$-ray bands. The middle and lower panels show the optical PD and PA evolution, respectively. Two active phases can be easily identified in the light curves: one from $0.5\tau_{lc}$ to $1.5\tau_{lc}$, the other from $2\tau_{lc}$ to $2.5\tau_{lc}$. An optical PA swing is present during the second active phase.}
\label{fig:temporal}
\end{figure}

\begin{figure}
\centering
\includegraphics[width=\linewidth]{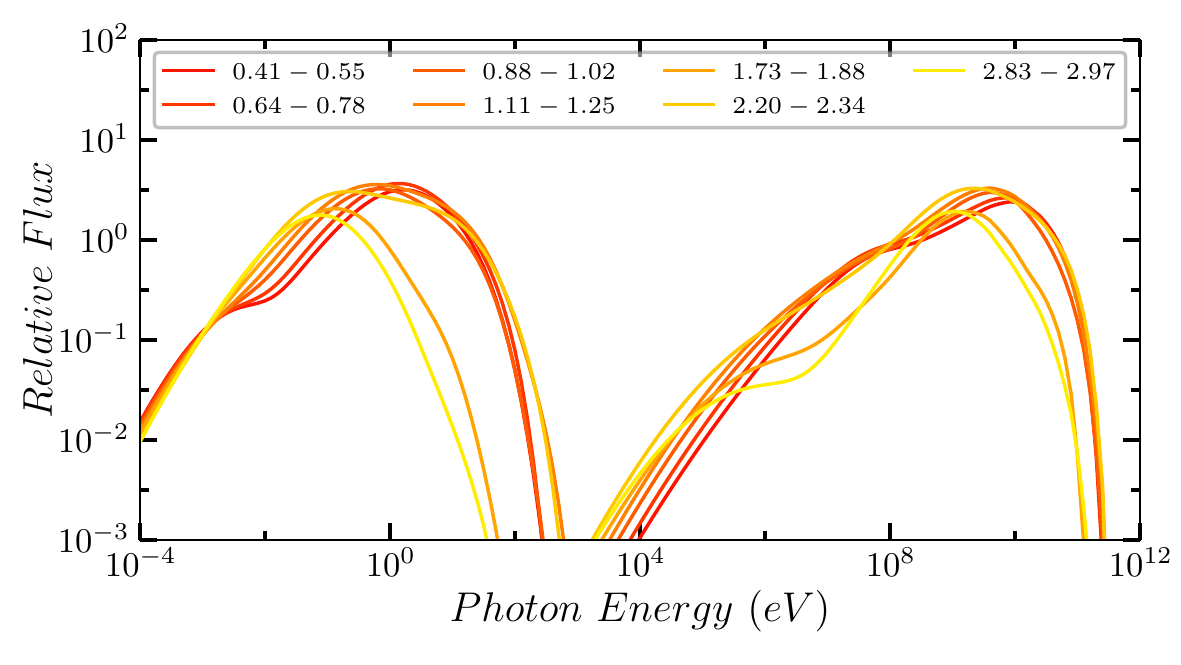}
\includegraphics[width=\linewidth]{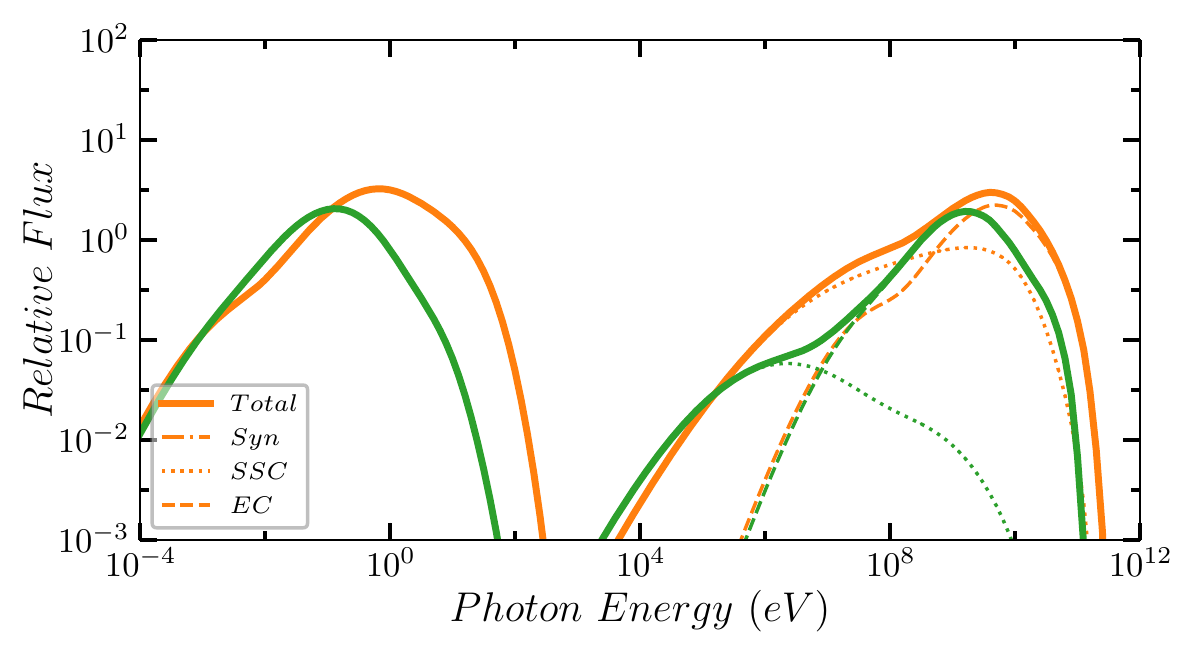}
\caption{Snapshots of the SEDs from the default simulation. Upper panel: the total SEDs in selected snapshots in units of the light crossing time $\tau_{lc}$: the first four snapshots cover the first active phase and the last three snapshots cover the second active phase in Figure \ref{fig:temporal}. Each snapshot integrates and averages over the same time span. Lower panel: the SEDs during the first active phase (orange) and the low state (green) between the two active phases. The total (solid), synchrotron (dash-dotted, overlapping with the low-energy component of the total SED), SSC (dotted), and EC (dashed) are explicitly shown for the active and low states.}
\label{fig:spectral}
\end{figure}

\begin{figure}
\centering
\includegraphics[width=\linewidth]{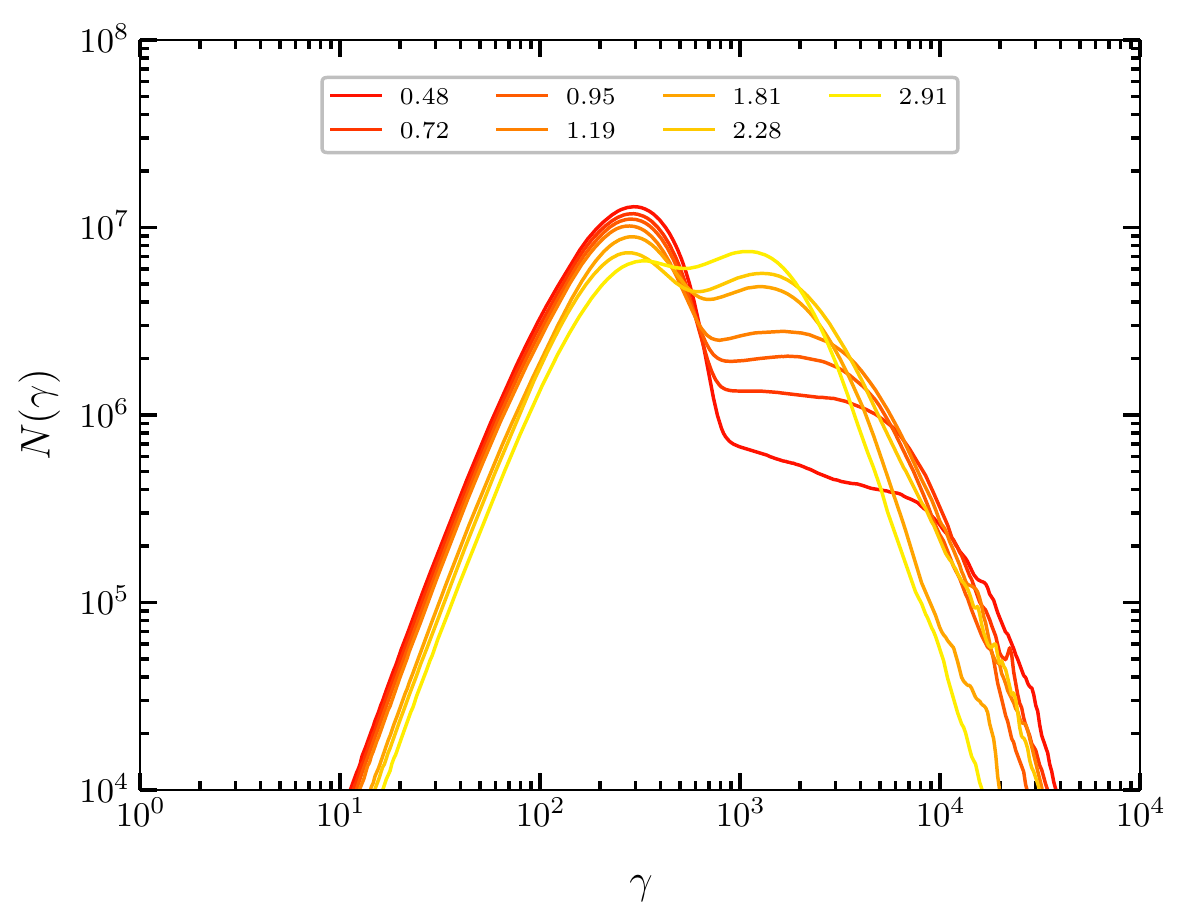}
\caption{Snapshots of the particle spectra from the default simulation. The particle spectra are selected at the middle of each epoch in the upper panel of Figure \ref{fig:spectral}.}
\label{fig:particle}
\end{figure}

In the default simulation setup, we assume that the optical, SSC, and EC components are comparable. Figure \ref{fig:temporal} shows the multi-wavelength light curves and optical polarization signatures. There are two active phases in the light curves, one from $0.5 \tau_{lc}$ to $1.5 \tau_{lc}$ and the other from $2\tau_{lc}$ to $2.5 \tau_{lc}$. There is an optical PA swing during the second active phase. Figures \ref{fig:spectral} and \ref{fig:particle} show the SEDs and particle spectra for selected epochs during the two active phases, respectively.

In a considerably magnetized environment with rather small guide field, after the initial perturbation the reconnection region quickly dissipates magnetic energy and accelerates particles into power-law distributions. As shown in Figure \ref{fig:particle}, the particle spectra are rather flat before the cooling break, so that the post-reconnection mean Lorentz factor is on average at $\gamma_c\sim 10^4$ within $\tau_{lc}$. The exact location of this cooling break, however, changes with time. Since the available magnetic energy is fixed in the simulation domain, the acceleration gradually wanes down. Therefore, at later time, the radiative cooling will push the cooling break towards lower energies, which is clearly shown in Figure \ref{fig:particle}. During the active phases, the low- and high-energy spectral components have similar flux levels (Figure \ref{fig:spectral} upper panel), due to the comparable cooling rate in our default setup. The high-energy spectral component is slightly wider than the synchrotron component though, since the peaks of SSC and EC do not exactly match. Given our parameters, during the active phases the SSC peaks at $E_{SSC}=E_{syn}\gamma^2 \delta \gtrsim \rm{GeV}$, where $E_{syn}$ is the synchrotron spectral peak, but the EC component can extend to $E_{EC}=E_{bb}\gamma^2\Gamma\delta \sim 10~\rm{GeV}$, where $E_{bb}$ is the peak of the black body spectrum (Figure \ref{fig:spectral} lower panel). Both SSC and EC are within the Thomson regime, thus we do not consider any Klein-Nishina effects in our simulations. The synchrotron, SSC and EC components follow a harder-when-brighter trend, which originates from the harder particle spectra during active phases as shown in Figure \ref{fig:particle}. Finally, the two snapshots at low states in the multi-wavelength spectra show much lower SSC flux than the EC ($1.73-1.88$ and $2.83-2.97$ spectral curves in the Figure \ref{fig:spectral} upper panel).

\begin{figure*}
\centering
\includegraphics[width=\linewidth]{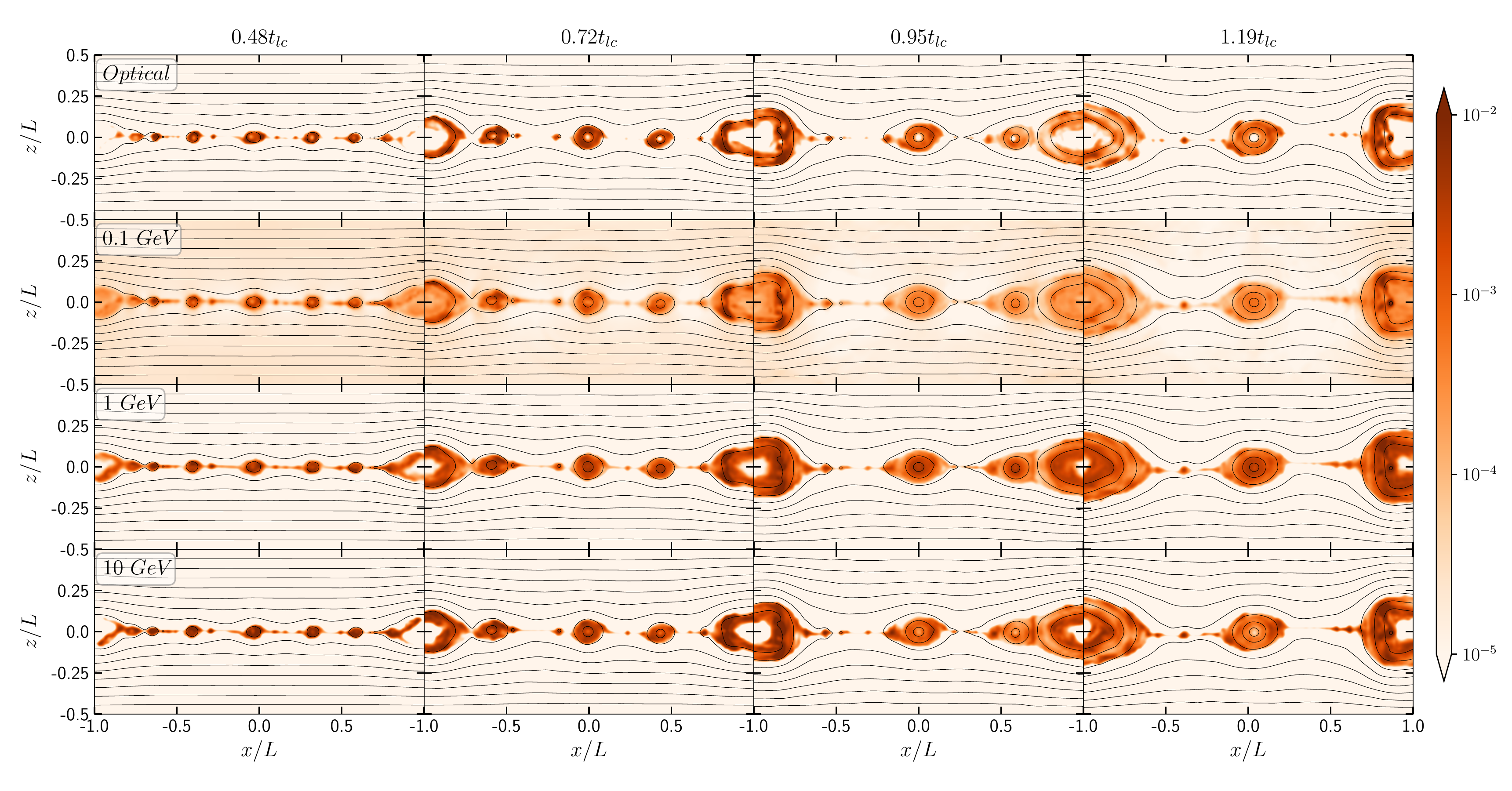}
\caption{From left to right columns are four snapshots of the optical (first row), $0.1~\rm{GeV}$ (second row), $1~\rm{GeV}$ (third row), and $10~\rm{GeV}$ (fourth row) emission maps.}
\label{fig:opticalgammamap}
\end{figure*}

Figure \ref{fig:opticalgammamap} shows four snapshots in the first active phase. Obviously, for both synchrotron and high-energy radiation, the emission concentrates in the plasmoids. However, the radiation is not uniform inside the plasmoids. It is clear that the synchrotron emission is stronger towards the edge of the plasmoids (Figure \ref{fig:opticalgammamap} first row). At the center, where the magnetic field is often the strongest in the plasmoids as shown in some previous works \citep{Zhang2018}, the synchrotron emission is trivial. \citetalias{Zhang2020} has discussed this phenomenon, which can be attributed to the fact that a considerable amount of particles are accelerated towards the edge of the plasmoids due to mergers. On the other hand, the $\gamma$-ray emission maps look different. They have not only the concentration of emission in similar regions of the synchrotron maps, but also exhibit a uniform component of Compton scattering. One obvious reason is that the $\gamma$-ray emission maps cover a much larger energy range, and for emission at $0.1-1~\rm{GeV}$, there is a considerable contribution from lower-energy electrons at energies below the cooling break. In particular, Figure \ref{fig:opticalgammamap} (second row) shows that the $0.1~\rm{GeV}$ emission has a small contribution from the upstream thermal electrons (towards the tail of the upstream thermal electrons, they have a Lorentz factor of $\gamma \sim 10^3$, which results in EC photons at $E_{EC}=E_{bb}\gamma^2\Gamma\delta \sim 0.1~\rm{GeV}$. \citetalias{Zhang2020} has thoroughly discussed that the electrons below the cooling break can last for a long time in the reconnection region, thus they can cover a much larger region (often the entire plasmoid) than those electrons at or beyond the cooling break. Since the seed photons for the EC are uniform in the emission region, these electrons then exhibit a uniform Compton scattering field in all plasmoids. This is also supported by the fact that the $\gamma$-ray emission maps appear more inhomogeneous and better resemble the optical emission map at $10~\rm{GeV}$.

\begin{figure}
\centering
\includegraphics[width=\linewidth]{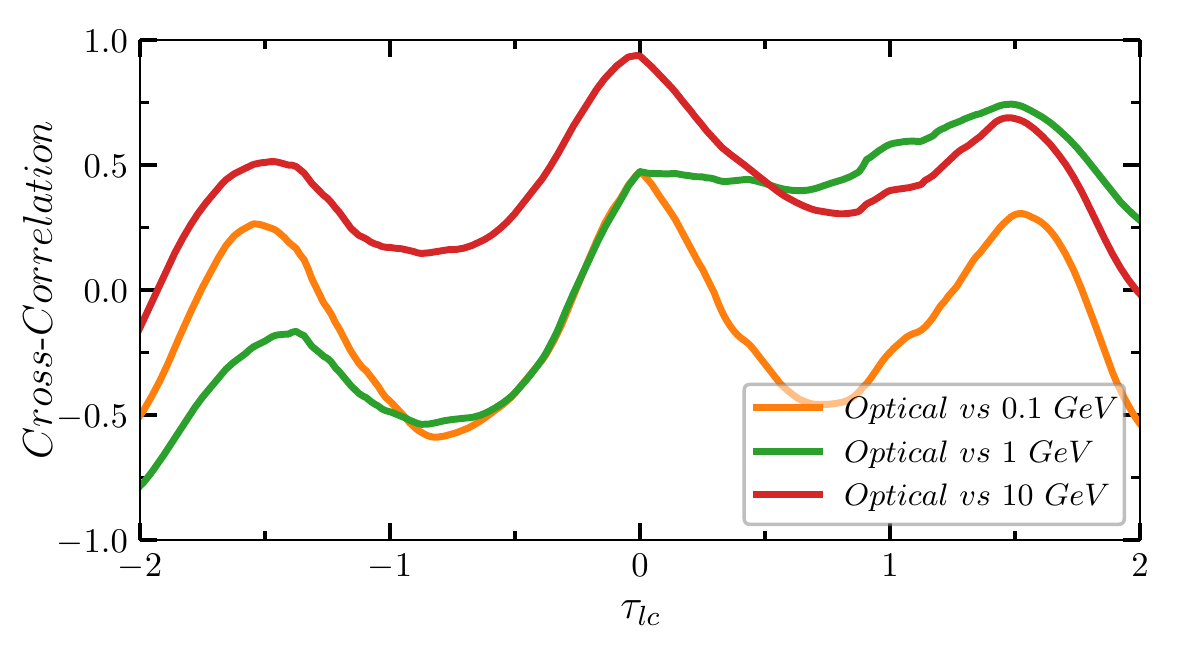}
\caption{Cross-correlation between the optical and $\gamma$-rays.}
\label{fig:correlation}
\end{figure}

During the first active phase, both the optical and $\gamma$-ray light curves exhibit several flares. While the first flare that peaks around $0.75\tau_{lc}$ is mostly due to the particle acceleration from the primary reconnection, the later peaks result from a couple of moderate plasmoid merger events. The first active phase ends when the primary reconnection is mostly saturated. The second active phase results from a large plasmoid merging into the plasmoid at the periodic boundary. As expected for most leptonic models, the optical and $\gamma$-ray light curves have nearly zero time lag (Figure \ref{fig:correlation}). Interestingly, the $1~\rm{GeV}$ light curve shows a very different cross-correlation function compared to the other two $\gamma$-ray bands. This is because this energy band marks the transition from SSC to EC. As we can see in Figure \ref{fig:spectral} (lower panel), during the flare the SSC contribution dominates the $0.1~\rm{GeV}$ band while the EC contribution dominates the $10~\rm{GeV}$ band. Since the rising and fading phases of synchrotron, SSC, and EC originate from the same particle distribution, the optical, $0.1~\rm{GeV}$ and $10~\rm{GeV}$ light curves are well correlated. However, both SSC and EC contribute to the $1~\rm{GeV}$ band. When the primary reconnection saturates and particles gradually cool down, the SSC contribution drops in the $1~\rm{GeV}$ band, but the peak of the EC component also moves to lower energy (Figure \ref{fig:spectral} lower panel green curve), which enhances the EC contribution in the $1~\rm{GeV}$ band. Consequently, the $1~\rm{GeV}$ flux does not drop much during the low state (Figure \ref{fig:temporal} upper panel). As a result, the $1~\rm{GeV}$ light curve does not show clear cross-correlation to the optical band compared to the other two $\gamma$-ray bands.

\begin{figure*}
\centering
\includegraphics[width=\linewidth]{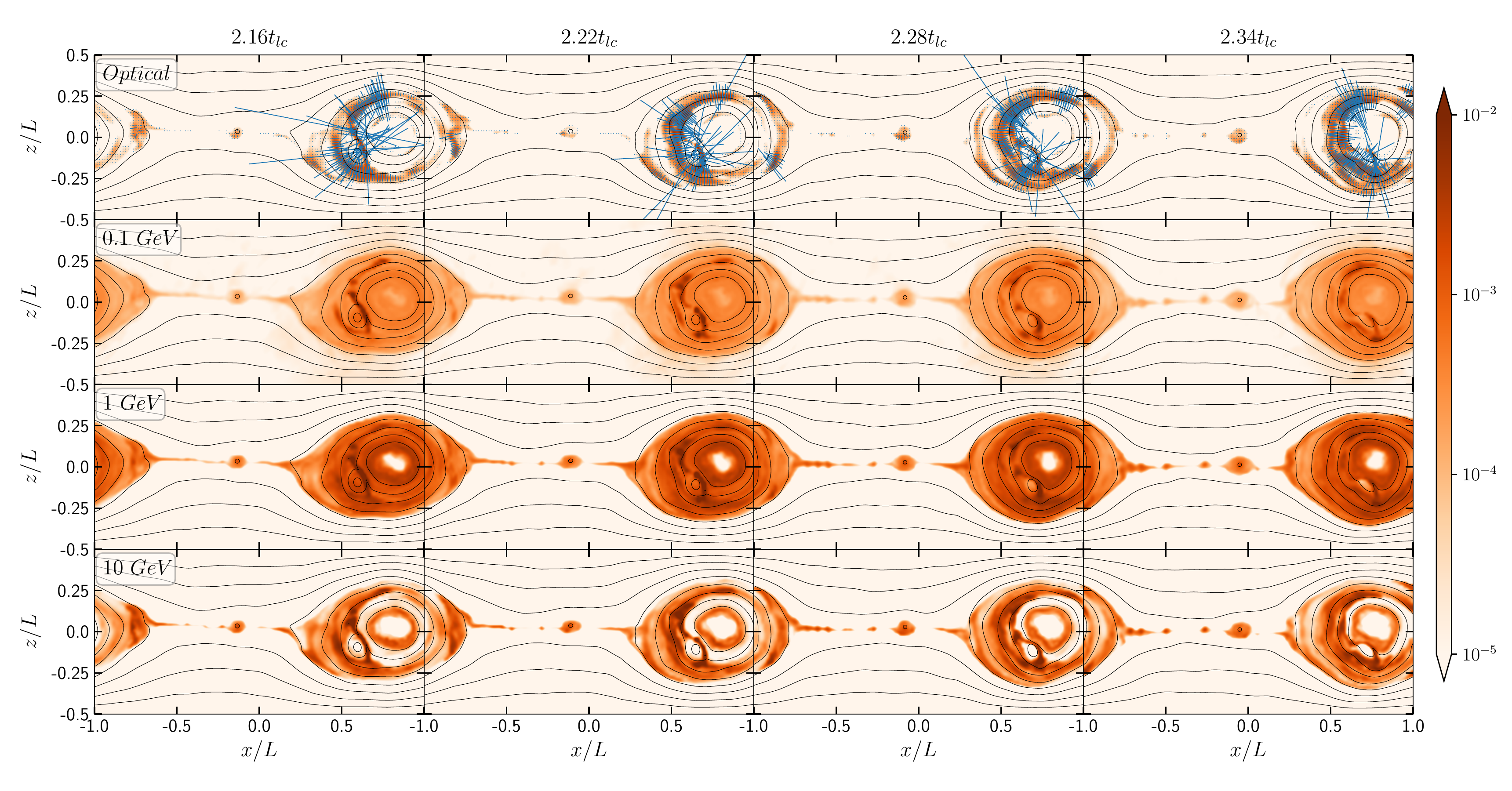}
\caption{Same as Figure \ref{fig:opticalgammamap}, but the four snapshots cover the optical PA rotation period as shown in Figure \ref{fig:temporal}. In the first row, the local relative polarized flux is plotted in blue lines, whose length represents the strength and direction represents the PA.}
\label{fig:PArotationmap}
\end{figure*}

Very interestingly, there is an optical PA swing during the second active phase (Figure \ref{fig:temporal} lower panel). Figure \ref{fig:PArotationmap} shows four snapshots during the swing. Similar to our previous works, the swing results from the newly accelerated particles at the merger site, which stream along the post-merger plasmoid. The polarized flux shown in Figure \ref{fig:PArotationmap} (first row) illustrates that the newly accelerated particles in the first snapshot at the interacting region (near the left edge of the plasmoid at the periodic boundary) strongly enhance the local polarized emission. Those particles then stream counterclockwise to the right side of the post-merger plasmoid in the next two snapshots. At the end of the swing (last snapshot), these particles are mostly cooled, so that the overall polarization direction goes back to the average polarization in the plasmoid, which happens to be the same as the beginning of the swing. This leads to the $180^{\circ}$ swing in Figure \ref{fig:temporal} (bottom panel). The small region that exhibits very strong local polarized flux in the optical band also appears as an enhanced Compton scattering region in all $\gamma$-ray bands. Additionally, the movement of this enhanced $\gamma$-ray region closely resembles the stream of the polarized emission region. This is expected, because the newly accelerated particles at the merger site that make the PA swing also contribute to Compton scattering. Since these particles occupy a small region in the post-merger plasmoid, they exhibit a very localized enhancement of $\gamma$-rays, which result in considerable flares. Therefore, the magnetic reconnection model predicts that the optical PA swings are simultaneous with $\gamma$-ray flares.

\subsection{Synchrotron Self Compton vs External Compton \label{sec:sscvsec}}

\begin{figure}
\centering
\includegraphics[width=\linewidth]{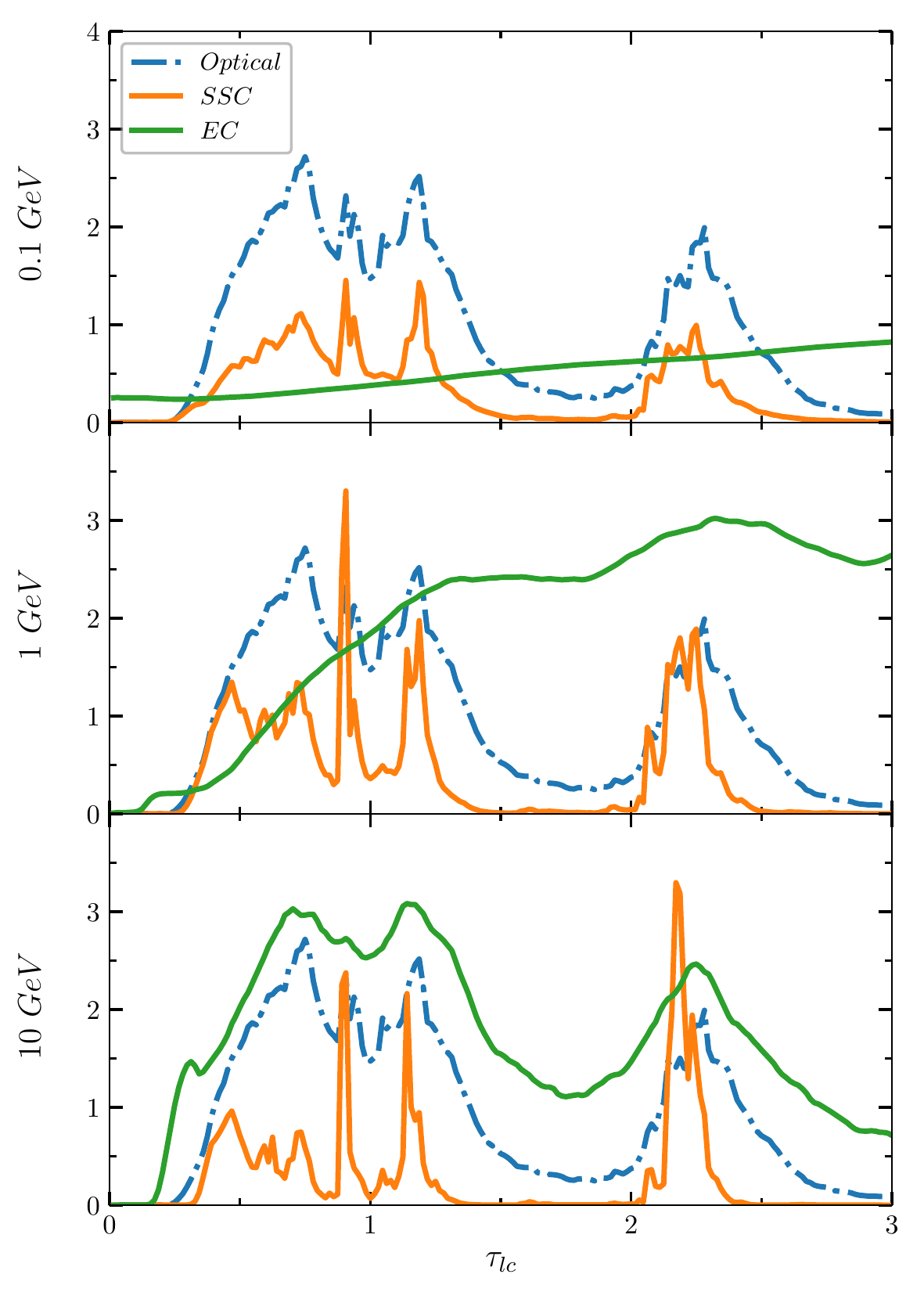}
\caption{SSC and EC light curves at the $0.1~\rm{GeV}$ (upper panel), $1~\rm{GeV}$ (middle panel), and $10~\rm{GeV}$ (lower panel) bands. The synchrotron light curve in the optical band is plotted in all panels for reference.}
\label{fig:sscvsec}
\end{figure}

\begin{figure}
\centering
\includegraphics[width=\linewidth]{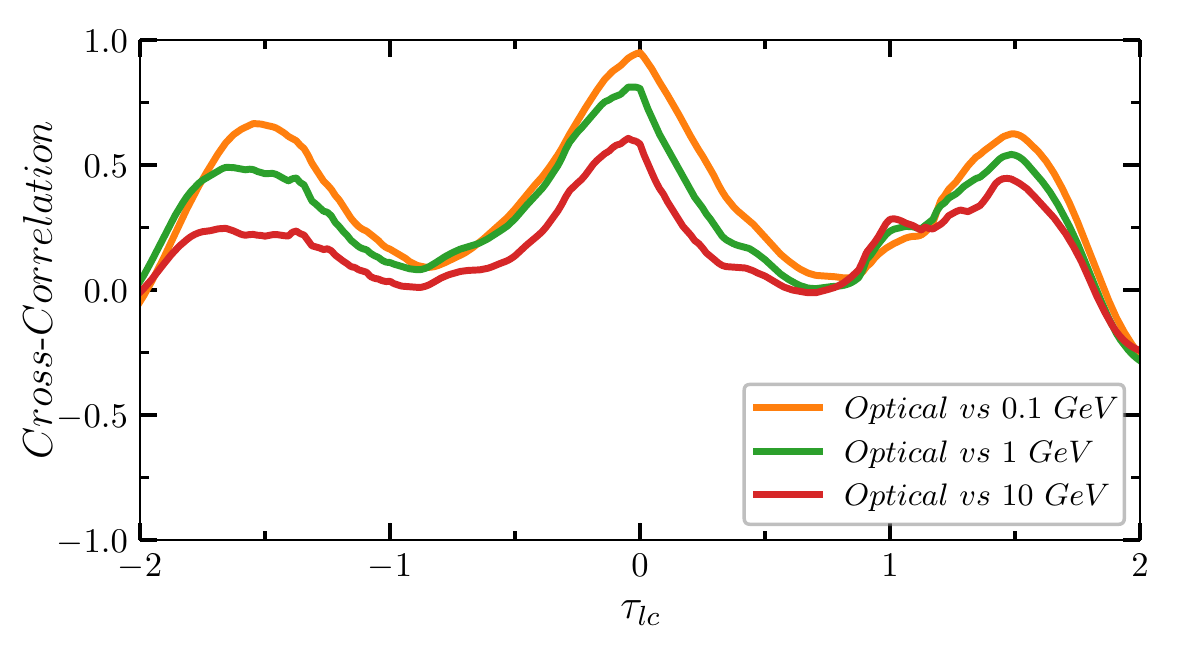}
\caption{Cross-correlation between the optical synchrotron emission and SSC.}
\label{fig:ssccorrelation}
\end{figure}

\begin{figure}
\centering
\includegraphics[width=\linewidth]{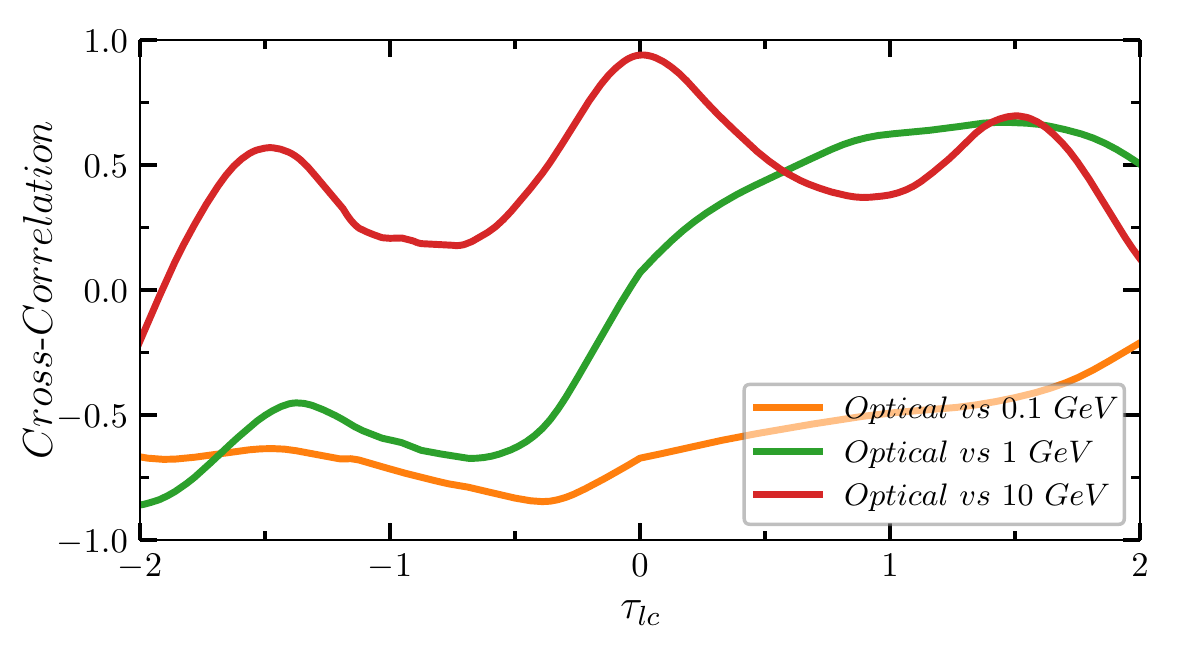}
\caption{Cross-correlation between the optical synchrotron emission and EC.}
\label{fig:eccorrelation}
\end{figure}

Figure \ref{fig:spectral} already shows that both SSC and EC have considerable contributions to the total $\gamma$-ray emission. It is therefore important to identify their respective contributions to multi-wavelength light curves and observable properties. Figure \ref{fig:sscvsec} shows the SSC and EC light curves in three $\gamma$-ray bands. Obviously, the EC emission does not show much variability or correlation with the optical band at low energies ($0.1-1~\rm{GeV}$, also see Figure \ref{fig:eccorrelation}), since the electrons that make low-energy EC $\gamma$-rays are of lower energies compared to those responsible for the optical synchrotron emission. At $10~\rm{GeV}$, the EC light curve nicely resembles the optical light curve, as the underlying electron energy is nearly the same. As we can see later in this section, it is generally true that strong particle acceleration often occurs where magnetic energy is actively dissipated. On the other hand, Figure \ref{fig:sscvsec} shows that the SSC light curves are correlated with the optical light curve. This is also evident by Figure \ref{fig:ssccorrelation}, which exhibits clear cross-correlation between the synchrotron and SSC emission. The reason is that the seed photons for the SSC are the entire synchrotron component. Since the nonthermal particles accelerated via magnetic reconnection have relatively hard spectra, as shown in Figure \ref{fig:spectral} (lower panel), in all $\gamma$-ray bands the SSC emission is mostly attributed to the high-energy electrons ($\gamma\sim 10^4$) that also produce the optical synchrotron emission. The different $\gamma$-ray photon energies are simply determined by the seed synchrotron photons, which can extend from infrared to optical frequencies. Therefore, even for three orders of magnitudes in $\gamma$-ray energies, the SSC light curves are expected to be well correlated with the optical emission.

\begin{figure}
\centering
\includegraphics[width=\linewidth]{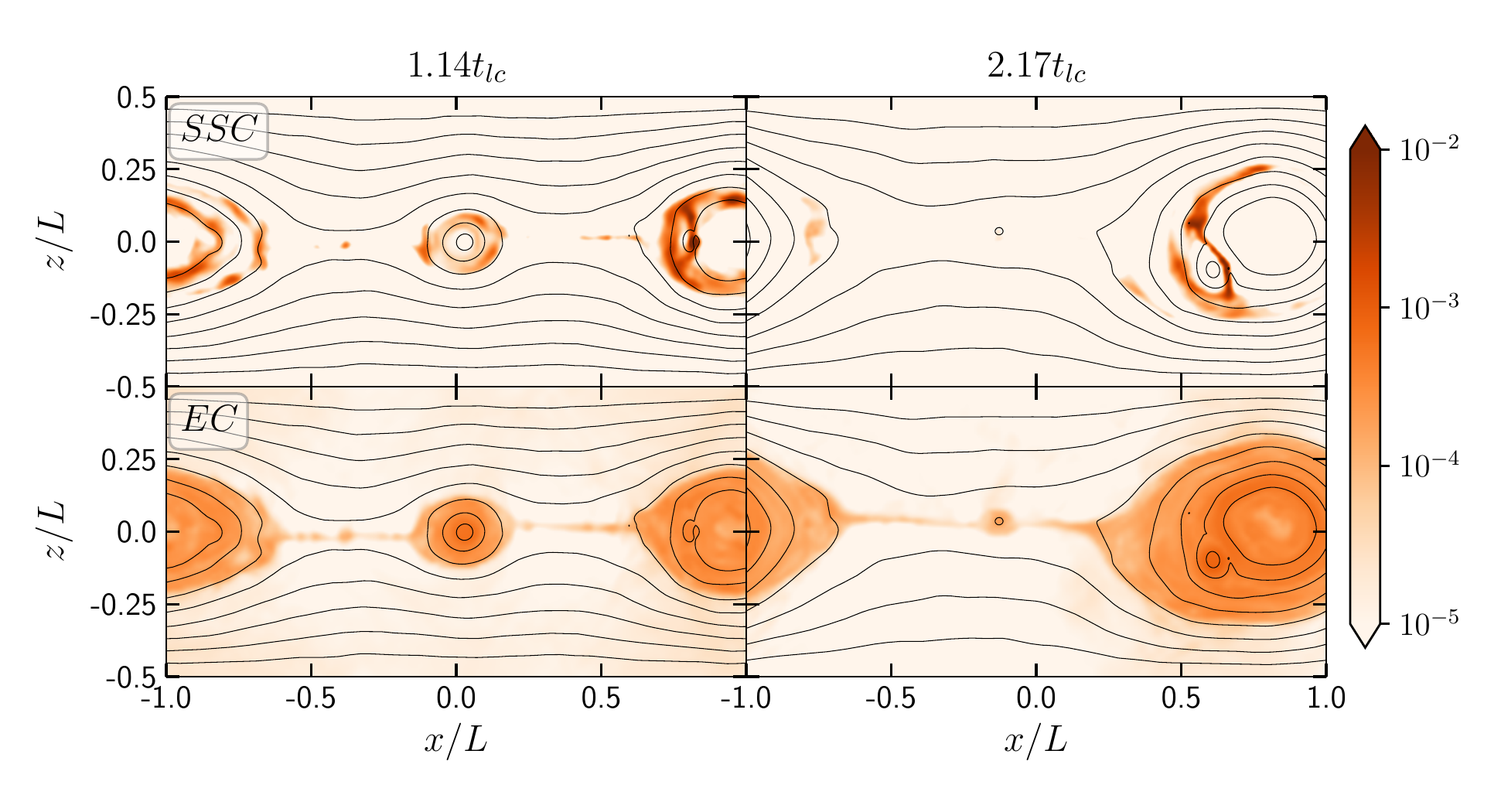}
\caption{Two snapshots of the SSC (top) and EC (bottom) emission maps at the $10~\rm{GeV}$ band. They are selected at second and third SSC peaks at this band.}
\label{fig:sscvsecmap}
\end{figure}

\begin{figure}
\centering
\includegraphics[width=\linewidth]{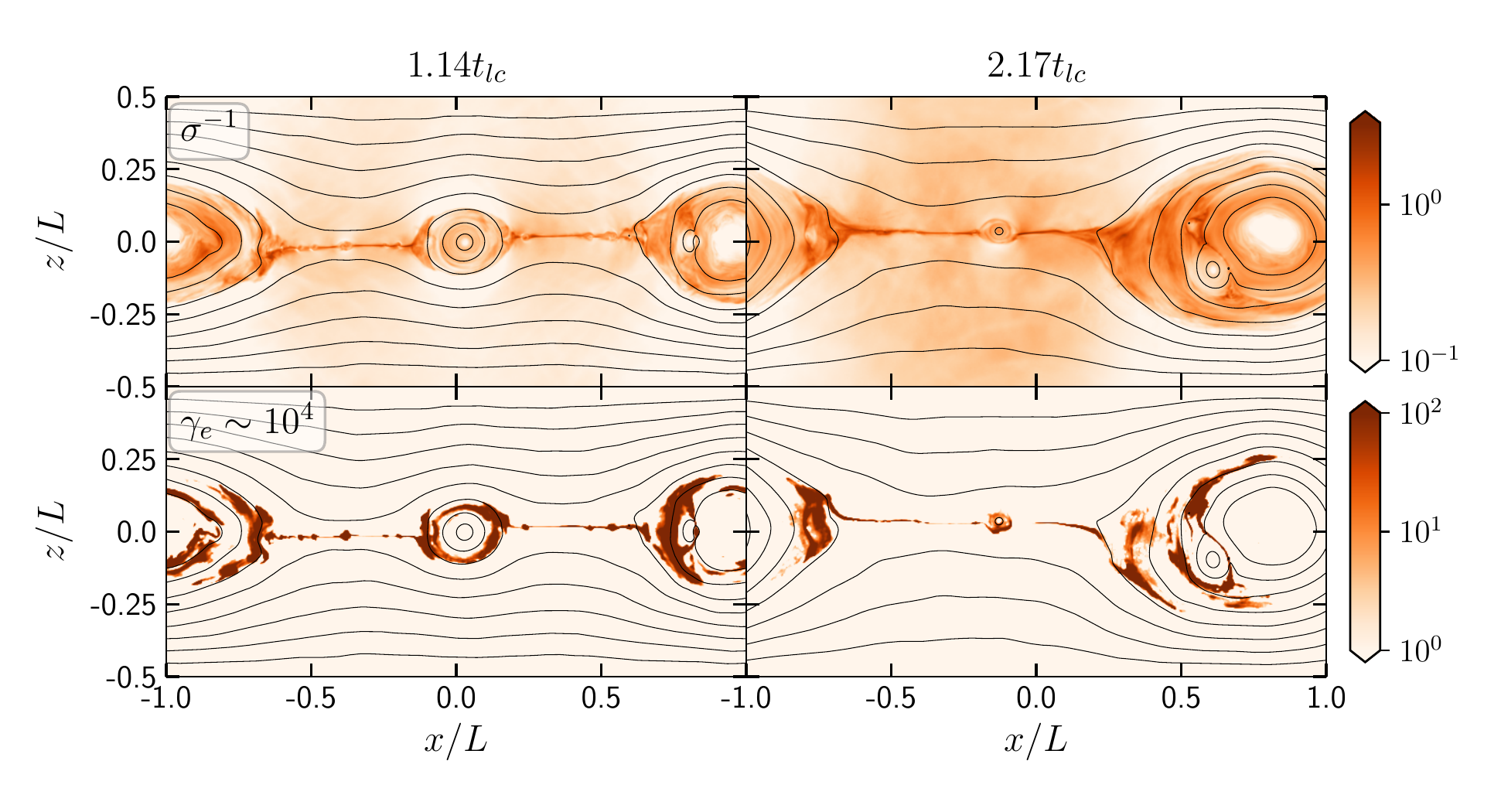}
\caption{Spatial distribution of the inverse of the local magnetization factor $\sigma^{-1}$ (top) and the high-energy ($10^4<\gamma<2\times 10^4$) electrons (bottom) at the same two snapshots in Figure \ref{fig:sscvsecmap}.}
\label{fig:phys}
\end{figure}

Very interestingly, the SSC light curves appear much more variable than the optical synchrotron and EC in all $\gamma$-ray bands. As clearly shown in Figure \ref{fig:sscvsec}, the SSC light curves feature very fast and large-amplitude flares during both active phases. However, by analyzing the local speed of the plasma, these fast variability patterns are not due to the local Doppler boosting effects often discussed in relativistic magnetic reconnection \citep[often referred to as the mini-jet model, see][]{Giannios2009,Sironi2016}. Instead, these signatures originate from the highly inhomogeneous distribution of high-energy particles in space. Figure \ref{fig:sscvsecmap} plots the spatial distribution of the $10~\rm{GeV}$ SSC and EC emissivity in the simulation domain. Apparently, unlike the EC contribution that is mostly uniform in all plasmoids, the SSC emission only occupies very localized regions at the edges of the plasmoids. Figure \ref{fig:phys} shows the spatial distribution of local magnetization factors and high-energy particles. The nonthermal particle distributions closely resemble the shape of the SSC emission map, indicating that the SSC emission is dominated by the most energetic electrons in the strongest acceleration regions, which are the merging sites of the plasmoids. This is also supported by the magnetization factor distribution, where we can observe that the regions with many high-energy particles have relatively low magnetization $\sigma<1$. This implies that in these very small regions, a considerable amount of the magnetic energy has been dissipated to accelerate particles. On the other hand, the EC emission includes a contribution from some lower-energy electrons, which are more uniformly distributed in the plasmoids. Therefore, its emission map covers a much larger region in the simulation domain and variability is less pronounced.

Based on the above physical picture, we can explain the fast and high-amplitude SSC flares. Since the SSC emission is dominated by the plasmoid merging sites, the duration of the flare will be on the order of the light crossing time of the interacting region of the plasmoid mergers, which is typically smaller than the radius of the plasmoid. Previous works suggested that the size of the plasmoids can be up to $\sim 10\%$ of the reconnection region \citep{Guo2015,Guo2016,Sironi2016,Petropoulou2016,Petropoulou2019,Loureiro2012}. Therefore, we expect that these fast flares should have time scales of approximately a few percent of the light crossing time of the reconnection region. This is consistent with Figure \ref{fig:sscvsec}, where the duration of the fastest flares is typically $\lesssim 0.1~\tau_{lc}$. Given that the SSC emission is mostly concentrated in small regions, it supports our assumption that the SSC only happens in very localized regions where the internal light crossing delays are not important. For the flare amplitude, two parameters are most important in the reconnection region, i.e., the magnetic energy density $u_B$ and the nonthermal particle energy $u_{non}$. The latter is essentially a portion of the dissipated $u_B$ during reconnection. For regions with efficient magnetic energy dissipation due to very fast magnetic reconnection (now $\sigma <1$), $u_B$ considerably drops and $u_{non}$ increases significantly, so that both synchrotron and EC flux increase (EC increases more than the synchrotron as the latter is also proportional to $u_B$). For the SSC, however, since both the seed photon density from the synchrotron and $u_{non}$ increase, its flare amplitude is further boosted to much higher level. Nevertheless, for regions with concentration of high-energy particles but considerable magnetization ($\sigma\gtrsim 1$), synchrotron cooling is still dominating and the SSC emission is thus suppressed. Consequently, the SSC emission is only strong in very localized regions with both low $\sigma$ and high nonthermal particle density, which occupy smaller area than the synchrotron emission map in the optical band (Figure \ref{fig:opticalgammamap} first row and Figure \ref{fig:sscvsecmap}).

\subsection{Synchrotron Cooling vs Inverse Compton Cooling \label{sec:cooling}}

As shown above, the SSC cooling is similar to the synchrotron cooling in that it only affects localized regions. On the other hand, both SSC and EC cooling effects are proportional to the seed photon density. It is therefore important to see how the radiative cooling due to the synchrotron and Compton scattering may affect the reconnection differently. To this end, we set up five different PIC simulations, where we set $r_e$ and $\mathcal{U}_{\star}$ parameters in the radiation reaction force to change the ratio between the synchrotron cooling and Compton scattering cooling, but the total cooling remains similar. Of note, since the Compton scattering cooling in the radiation reaction force is treated as a uniform cooling due to a preexisting photon field $\mathcal{U}_{\star}$, while the synchrotron cooling is calculated based on the local electromagnetic field on the fly, it is not possible to keep the total cooling in every cell exactly the same. Instead, we take the average cooling break in the total particle spectrum as an indicator of the total cooling in the simulation domain. Additionally, if the synchrotron cooling changes, its emissivity obviously follows the change, which can affect the SSC emission. As shown above, the SSC light curves are significantly different from EC light curves, thus the final $\gamma$-ray light curves can appear very different due to the ratio between SSC and EC rather than the cooling effects. We choose to fix the average ratio between SSC and EC as in the default setup in Section \ref{sec:opticalgamma} so as to ensure apples-to-apples comparison.

\begin{figure}
\centering
\includegraphics[width=\linewidth]{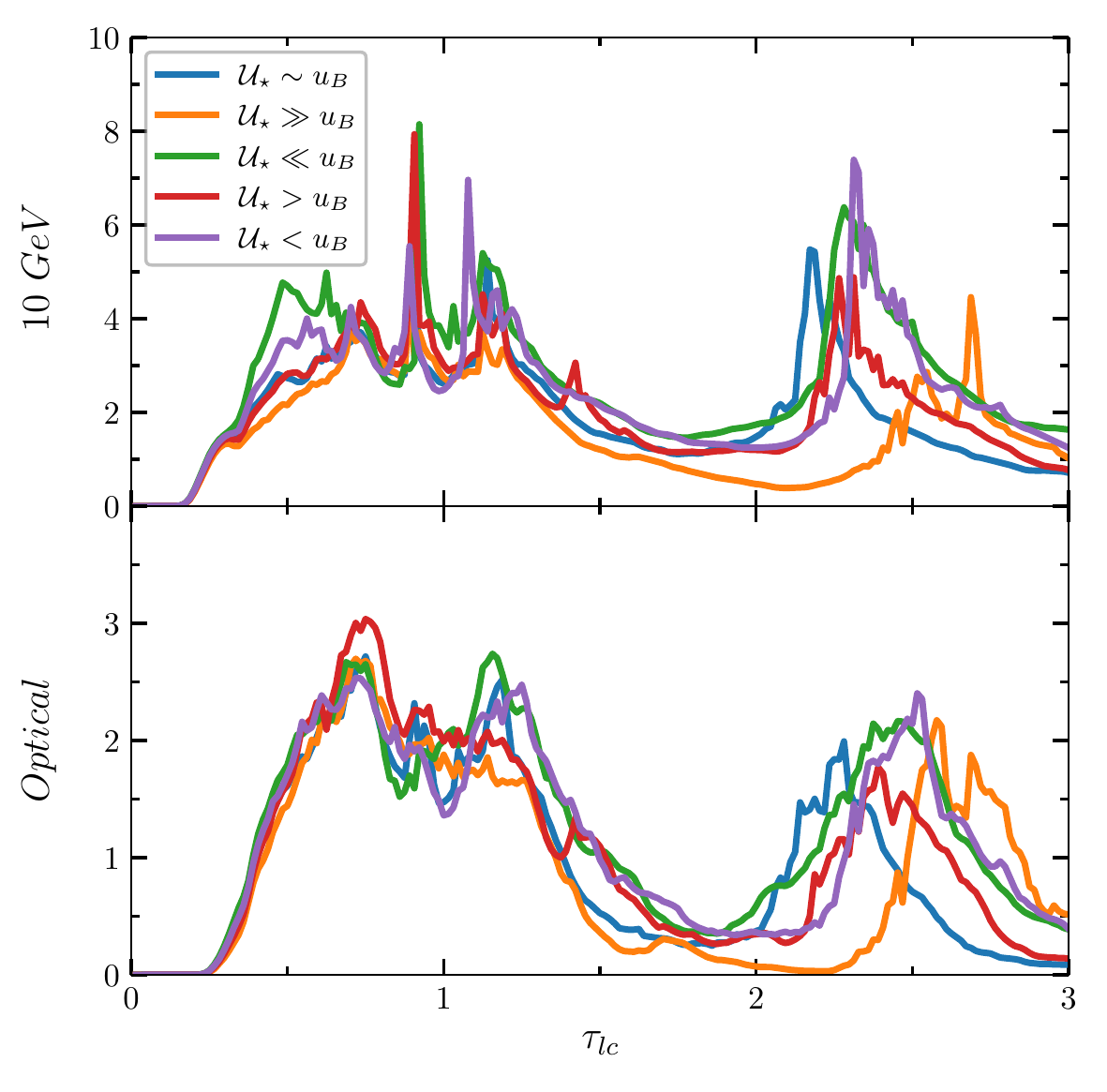}
\caption{$10~\rm{GeV}$ $\gamma$-ray and optical light curves for different ratios between the synchrotron and inverse Compton cooling in the PIC simulation. The light curves are normalized to similar flux to be shown in the same plot.}
\label{fig:cooling}
\end{figure}

Figure \ref{fig:cooling} shows the result. We can see that the light curves in the first active phases are very similar for any ratio between synchrotron and Compton scattering cooling. The small differences are likely due to the fact that the total cooling is not exactly the same in every cell. The second active phases show some time lags between different runs. However, we have examined the emission map and noticed that the plasmoid mergers for different runs do not happen at the same time step and location. Thus, the time lags can be attributed to the differences in the plasmoid mergers. As shown in previous works \citep{Guo2015,Guo2016,Werner2018,Sironi2016,Petropoulou2016,Zhang2018}, plasmoid mergers are generally random events even under periodic boundary conditions. Therefore, we conclude that the synchrotron and Compton scattering cooling terms in the radiation reaction force have similar effects on the radiation signatures. In this situation, the SSC cooling can be considered as an additional cooling contribution on top of the local synchrotron cooling. Even though this may have some impact on the local particle spectra, since the SSC emission only occupies very small regions and flashes in very short time, we believe that its effects will not significantly alter the overall radiation signatures.

\section{Implications for Observations \label{sec:observation}}

Blazars have been observed for decades. In particular since the launch of {\it Fermi}, multi-wavelength blazar monitoring programs have matured, resulting in many interesting systematic patterns in radiation signatures. Unlike individual flaring events, these behaviors may reveal general physical conditions and processes in the blazar emission region, which should be studied self-consistently with minimal free model parameters. Relativistic magnetic reconnection is a primary plasma process that can efficiently dissipate magnetic energy in a magnetic-driven blazar jet. For the first time, we investigate the multi-wavelength radiation and polarization signatures arising from reconnection in a pre-existing current sheet in the blazar emission environment, directly based on combined PIC and polarized radiation transfer simulations. Our approach therefore involves a minimal number of free parameters, while it can explore all correlated observable signatures under a consistent physical picture. Here we discuss several interesting signatures from our leptonic blazar model.

A harder-when-brighter trend is explicitly shown in both the optical and {\it Fermi-LAT} $> \rm{GeV}$ $\gamma$-ray band based on our simulations. This trend is frequently reported in observations \citep{Giommi90,Abdo10,Krauss16}. This feature results from the particle spectra. However, we find that this trend does not apply to the entire high-energy spectral component. The reason is that the high-energy spectral component may consist of multiple radiation contributions in blazars. In our simulation, both SSC and EC contribute to the high-energy emission. Although the two components are comparable during the high states, in the low states the two can have very different flux levels. As a result, in the high state the $\lesssim 0.1~\rm{GeV}$ $\gamma$-ray band shows a rather flat spectrum dominated by the SSC contribution, but in the low state it exhibits a rising spectrum dominated by the EC contribution (see Figure \ref{fig:spectral}). This issue generally applies to any blazar spectral modeling with multiple radiation processes in the high-energy spectral component, including the lepto-hadronic models \citep{Boettcher2013,Cerruti2019,Zhang2019,Keivani2018}, which can be tested with future MeV $\gamma$-ray telescopes such as {\it AMEGO} \citep{McEnery19,Rani2019b}. We note that our setup targets at a flat-spectrum radio quasar that peaks in the GeV band, and the harder-when-brighter trend is clear around or beyond the peak. For blazars that peak at other energies, the harder-when-brighter trend in the high-energy component should show up around and beyond the respective $\gamma$-ray peak.

We find that the optical and $\gamma$-ray light curves are correlated under the leptonic scenario. This is because both synchrotron and Compton scattering from magnetic reconnection are due to newly accelerated particles in the primary reconnection region and/or plasmoid mergers. Due to the rather hard particle spectra from reconnection, high-energy particles have significant contribution to the $\gamma$-ray flux via Compton scattering. Since the optical synchrotron emission is also dominated by high-energy particles, the two radiation processes then originate from the same particles. Although the PIC simulation domain is too small to see proton acceleration to very high energies, we find that the protons also have higher energies in the same region where electrons receive significant acceleration. Previous papers have shown that both protons and electrons can be efficiently accelerated in reconnection \citep{Guo2016}, thus we expect that optical and $\gamma$-ray light curves should be closely correlated under the hadronic scenario as well. In either the pure leptonic or lepto-hadronic scenario, the energy band that marks the transition between two radiation mechanisms may show light curves that appear not well correlated with the optical and other $\gamma$-ray bands. This feature is potentially significant for flat-spectrum radio quasars, where the transition band locates somewhere in the MeV to GeV bands, which can be examined by {\it Fermi-LAT} and future MeV telescopes such as {\it AMEGO}. We note that the time delay between the low- and high-energy emission may not be exactly zero. This is due to the highly inhomogeneous distribution of magnetic fields and particles in the reconnection region. Although the synchrotron, SSC and EC emission regions in our simulations are approximately co-spatial, their sizes are not the same. In general, the SSC emission size is much smaller than that of the synchrotron, which is smaller than that of the EC emission. As a result, we observe that the SSC emission is the most variable emission with many spikes in the light curve, while the EC emission is not as variable as the synchrotron band. The combined SSC and EC emission in the $\gamma$-ray bands therefore may show minor time delay compared to the optical band.

Finally, the optical PA swings are clearly accompanied by $\gamma$-ray flares. As shown in previous works, all PA swings are associated with major plasmoid mergers \citep{Zhang2018,Zhang2021b,Hosking2020}. These mergers will always lead to strong particle acceleration, which result in significant SSC and EC emission. In particular, while the EC flare amplitude during major plasmoid mergers is comparable to that of the synchrotron flares, due to the highly inhomogeneous distribution of particles and magnetic fields, the SSC flare amplitude can be much larger (see Figure \ref{fig:sscvsec}). Therefore, in observations we expect a strong correlation between optical PA swings and $\gamma$-ray flares, which is consistent with observations \citep{Marscher2010,Larionov2013,Blinov2015,Blinov2018}.

\section{Summary and Discussion \label{sec:discussion}}

To summarize, we have studied the multi-wavelength radiation and optical polarization signatures from relativistic magnetic reconnection in the blazar emission environment, based on combined PIC and polarized radiation transfer simulations. The reconnection process starts from a pre-existing current layer with nearly anti-parallel magnetic field lines (guide field $B_g=0.2$) in a proton-electron plasma. Our choice of parameters is characteristic for a flat-spectrum radio quasar whose high-energy spectral component peaks in the {\it Fermi-LAT} $\gamma$-rays. The combined PIC and polarized radiation transfer simulations account for both synchrotron and Compton scattering under the leptonic scenario. We have reached interesting conclusions as follows:
\begin{enumerate}
\item A harder-when-brighter trend is present in both optical and {\it Fermi-LAT} bands.
\item The optical light curves are correlated with $\gamma$-rays with nearly zero time delays.
\item Optical PA swings are accompanied by multi-wavelength flares.
\item SSC is much more variable with larger flare amplitude than synchrotron and EC due to the highly inhomogeneous particle distribution in the reconnection region.
\item The interface of the plasmoid merger events can have strong magnetic energy dissipation and particle acceleration, where the magnetization factor can drop to $\sigma \lesssim 1$.
\item As long as the total radiation cooling is similar, synchrotron and Compton scattering cooling effects in the PIC simulations are not clearly distinguishable from radiation signatures.
\end{enumerate}

A key discovery is the effect of highly inhomogeneous particle distributions in space on the SSC emission. Our simulations show that the magnetization factor can drop below 1 within interacting regions of plasmoid mergers. As a result, the local magnetic energy density may be smaller than the particle energy density, so that the synchrotron emission is suppressed, but the SSC emission is enhanced. However, this interface can only survive for a short period of time and the newly accelerated particles in these regions are quickly cooled. The combined effects lead to the very fast and high amplitude SSC $\gamma$-ray flares as shown in Figure \ref{fig:sscvsec}. In the total $\gamma$-ray light curves, these flares appear as very bright spikes on top of the active phases, with flaring time scales $\sim 0.1\tau_{lc}$. They may explain the fast flares observed in $\gamma$-rays \citep{Aharonian2007,Albert2007,Ackermann2016}. Previous explanations of fast blazar flares typically involve mini-jets in reconnection, which are fast-moving plasmoids in the reconnection region \citep{Giannios2009,Guo2015,Sironi2014,Sironi2016,Christie2019,Nalewajko2018}. If the bulk motion of nonthermal particles is along our line of sight, all radiation processes experience Doppler boosting. The inhomogeneous particle distributions in reconnection can play a complementary role to mini-jets in producing fast high-energy flares. This feature, however, does not depend on the line of sight or bulk relativistic motion of particles, and mostly applies to the SSC emission. The inhomogeneous distributions, as shown in our simulations, do not necessarily happen on kinetic scales. Basically, collisions between two magnetic structures, such as the plasmoids in reconnection, will naturally lead to strong magnetic energy dissipation and particle acceleration if the interacting region forms a current sheet. These have already been suggested in previous MHD simulations, which can lead to multiple flares on top of an active epoch \citep{Deng2016}. Similar to the fast-moving plasmoids, in reconnection inhomogeneous particle distributions also favor a low guide field and considerably magnetized environment, which can lead to many plasmoid mergers. The inhomogeneity of particles is thus an important feature that naturally arises from reconnection, which should be further studied with both kinetic-scale and fluid-scale simulations.

The above peculiar features can have very interesting implications on orphan $\gamma$-ray flares and the recent neutrino blazar flare event as well \citep{Icecube2018,Abeysekara2017,Krawczynski2004,Blazejowski2005}. Under the leptonic scenario, since the interface of the plasmoid merger may have a nonthermal electron energy density larger than the magnetic energy density, then the SSC flux can be much higher than the synchrotron flux. We note that our simulation setup only considers a pre-existing current sheet. In reality, the blazar emission region may have much larger emission regions that contribute to some quiescent state flux. Due to the relatively low synchrotron flare amplitude, it is possible that the optical flare may not be observable due to the quiescent state flux from the large emission region. However, the SSC flare amplitude is much higher, which can appear as a flare on top of the quiescent state flux. This may lead to an orphan $\gamma$-ray flare, whose duration is about one tenth of the light crossing time scale of the current sheet. Such orphan flares only apply to SSC, but not EC emission. Under the hadronic scenario, the interface in the plasmoid mergers will have a concentration of newly accelerated protons and relatively low magnetic field. The fresh protons will produce neutrinos via photomeson processes with the local photon field, but the secondary synchrotron emission from pairs may be suppressed, resulting in neutrino emission with low X-ray secondary synchrotron emission. Nonetheless, since our simulations are limited to the kinetic scales, both the orphan $\gamma$-ray flares under the leptonic scenario and neutrino emission under the hadronic scenario need to be further examined in a large-scale, realistic blazar emission environment.

It is important to note that previous studies have found the Compton dominance, which is the ratio between the luminosity of the Compton scattering component over the synchrotron component, is typically less than 0.1 \citep[][however, see \citealt{Christie2020}, which suggests that local Doppler boosting can enhance the Compton dominance]{Morris2019,Christie2019}. We note that these previous works generally assumed that the synchrotron photons are distributed over the entire plasmoid. In our simulations \citepalias[and see also][and \citealt{Hosking2020}]{Zhang2020}, synchrotron photons are clearly not uniform in the plasmoid. Particularly during plasmoid mergers, which are responsible for the strong SSC flares in our simulations, the majority of synchrotron emission is concentrated in the thin layer of the plasmoid merging site. This region typically occupies only a few to ten percent of the merging plasmoids. Following the calculation in \citet{Morris2019} for instance, the synchrotron photon energy density during plasmoid mergers will be higher than their estimates by at least one order, due to the highly inhomogeneous synchrotron emission. Since the SSC luminosity is proportional to the synchrotron photon energy density, this means that the Compton dominance is comparable to or larger than 1. If this inhomogeneity in kinetic-scale simulations can be extrapolated to a realistic blazar zone environment, the fast SSC flares in our results should be observable.

Generally speaking, the multi-wavelength blazar spectra and spectral evolution are similar for relativistic magnetic reconnection and shocks \citep{Baring2017,Boettcher2019b}. However, our simulations suggest that reconnection predicts highly variable $\gamma$-ray light curves including some fast flares, while the light curves are typically smooth under the shock scenario. Another key diagnostic is the optical polarization. As shown in our simulations, the optical polarization signatures are highly variable during $\gamma$-ray flares, including $\gtrsim 180^{\circ}$ PA swings. Although shocks and turbulence may also lead to PA swings \citep{Marscher2014,Boettcher2019b}, reconnection also predicts fast and large-amplitude polarization variations on the order of $\sim 0.1\tau_{lc}$. For typical blazar flares on the order of several hours to days, the fast polarization variations in reconnection are likely on the order of sub-hour to a few hours, which require high cadence optical polarization monitoring.

We want to point out two caveats in our simulations. One is that our choice of parameters and periodic boundary settings leads to no relativistic motion of plasmoids or a bulk of nonthermal particles. Not only does this ignore all mini-jet boosting in the multi-wavelength light curves, but it also eliminates local enhancements of synchrotron photon fields due to Doppler boosting. Right now our simulations are performed based on the assumption that SSC is only strong in very localized regions (which is true due to the inhomogeneous particle distributions), so that the internal light crossing effects are not important. However, if mini-jets are present, this assumption may break down, which can bring additional complexity in SSC processes \citep[see, e.g.,][for a complete description of internal light crossing effects on SSC]{Chen2012,Chen2014}. The other issue is that our simulations are 2D. It is already known that reconnection in 3D can lead to strong turbulence \citep{Guo2020b,Werner2021,Li2019}. The turbulence can have a strong impact on the plasmoid mergers and the inhomogeneity of particle distributions. These caveats and their effects should be explored with future 3D simulations.

\acknowledgments{We thank the anonymous referee for very helpful and constructive comments. H.Z. and F.G. acknowledge support from DOE grant DE-SC0020219. H.Z. and D.G. acknowledge support from NSF AST-1910451. The work by X. L. is funded by the National Science Foundation grant PHY-1902867 through the NSF/DOE Partnership in Basic Plasma Science and Engineering and NASA MMS 80NSSC18K0289. D.G. also acknowledges support from the NASA ATP NNX17AG21G and NSF AST-1816136 grants. F.G. also acknowledge the support from DOE through OFES and the LDRD program at LANL, and NASA ATP program through grant NNH17AE68I. The work of M. B. is supported by the South African Research Chairs Initiative (grant no. 64789) of the Department of Science and Innovation and the National Research Foundation\footnote{Any opinion, finding and conclusion or recommendation in this material is that of the authors and the NRF does not accept  any liability in this regard.} of South Africa. T.L. is supported by the NASA Postdoctoral Program at Goddard Space Flight Center, administered by USRA.  Simulations and analysis were performed at Texas Advanced Computing Center (TACC), National Energy Research Scientific Computing Center (NERSC) and LANL institutional computing.}

\bibliography{PIC+Pol3}
\bibliographystyle{aasjournal}



\end{document}